\documentclass[aps,prd,twocolumn,nofootinbib,showpacs]{revtex4}

\usepackage{graphicx}
\usepackage{dcolumn}
\usepackage{bm}
\usepackage{epsfig}

\begin{document}


\title{Light Quark Resonances in $\bar{p} p$ Annihilations at $5.2$ GeV/c}

%
\author{I.~Uman}
\author{D.~Joffe\footnote{Now at: Department of Physics, Southern Methodist University, Dallas, Texas 75275}}
\author{Z.~Metreveli}
\author{K.~K.~Seth}
\author{A.~Tomaradze}
\author{P.~Zweber}
\affiliation{Northwestern University, Evanston, Illinois 60208}


\begin{abstract} 
Data from the Fermilab E835 experiment have been used to study 
the reaction $\bar p p \rightarrow \eta\eta\pi^0$ at 5.2 $GeV/c$.
A sample of 22 million six photons events has been analyzed to construct
the Dalitz plot containing $\sim80k$ $\eta\eta\pi^0$  events.
A partial wave analysis of the data has been done.
Six $f_J$-states decaying into $\eta\eta$  
and five $a_J$-states decaying into $\eta\pi^0$ are identified 
in the mass region $\sim1.3$ and 2.4 $GeV$, and their 
masses, widths and spins are determined by 
maximum likelihood analysis of the data.
Two $f_0$ states are identified with the popular candidates for the 
lightest scalar glueball, $f_0(1500)$ and $f_0(1710)$.
\end{abstract}

\maketitle

\section{Introduction}

Antiproton-proton annihilations leading to final states containing 
light-quark structures constitute very attractive and powerful tools
for the study  of the spectroscopy of $q\bar q$ mesons, glueballs, 
and hybrids.
Indeed, major contributions to  light-quark spectroscopy have been made 
by the study of $\bar pp$ interactions with stopped and low energy 
antiprotons \cite{Amsler:1997up}.
In this paper we report a study of the isoscalar and isovector light quark resonances
populated in the reaction $\bar p p \rightarrow \eta\eta\pi^0$
in the collision of 5.2 GeV/c antiprotons with protons at rest.
The reaction is very selective. It can only populate $q\bar q$ states
with $J^{PC}=even^{++}$, isoscalar, or $f$-states decaying into $\eta\eta$, 
and isovector, or $a$-states decaying into
$\eta\pi^0$. The $\eta\eta$ states may contain admixtures of low lying 
$0^{++}$ and $2^{++}$ glueballs, and the $\eta\pi^0$ states may include 
states with non-$q\bar q$ $J^{PC}$, such as the $1^{-+}$ hybrids.

In the spectroscopy of light-quark mesons there remain many open questions.
Theoretical calculations  \cite{Godfrey:xj} for states with angular momentum $L\le4$  predict twelve $f_J$-states, 
and six $a_J$-states in the mass range 1 GeV-2.5 GeV.
In addition, two glueball states, with $I=0$, $J^{PC}=0^{++},2^{++}$, and 
at least one manifestly exotic $q\bar qg$ hybrid with  $I=1$, $J^{PC}=1^{-+}$ 
are predicted in this mass range.
In different experiments many $f_0$ and $f_2$ states 
have been claimed, but PDG'04 \cite{Eidelman:2004wy} considers only 
four  $f_0$ states, four $f_2$ states, and five $a_{0,2,4}$ 
to be established well enough  to be included in the meson summary list. 
In the present investigation we present new determinations of masses and 
$J^{PC}$ of several of the established $f_J$ and $a_J$ states, 
and provide evidence for several previously unconfirmed states.

\section{Data and Event Selection}

In the present analysis data from Fermilab $\bar p p$ annihilation 
experiment E835 is used. The stored and cooled antiprotons circulating 
in the Fermilab Antiproton Accumulator
intersect a hydrogen cluster jet gas target, and the reaction 
products are detected in a detector system which surrounds the 
interaction region. 
Full details of the detector and its performance can be found elsewhere  
\cite{Garzoglio:2004kw}. 
Photons are detected in the lead glass central detector in  
the angular range $10.6^{o}<\theta <70.0^{o}$ and $2\pi$ in $\phi $, 
with an energy resolution of 
$\frac{\sigma_E(E)}{E}=\frac{0.06}{\sqrt{E(GeV))}}+0.04$.
Luminosity was measured by detecting recoil protons
at $\sim87.5^{o}$. The data were collected at $\bar p$ 
momenta of 5.2 GeV/c corresponding to center of mass energy range of 
$E_{cms} = 3.409-3.418 $ GeV with a total luminosity of 
$11.4 pb^{-1}$. 

\begin{figure*}[tb!]
\setlength{\unitlength}{1cm}
\includegraphics[height=.23\textheight]{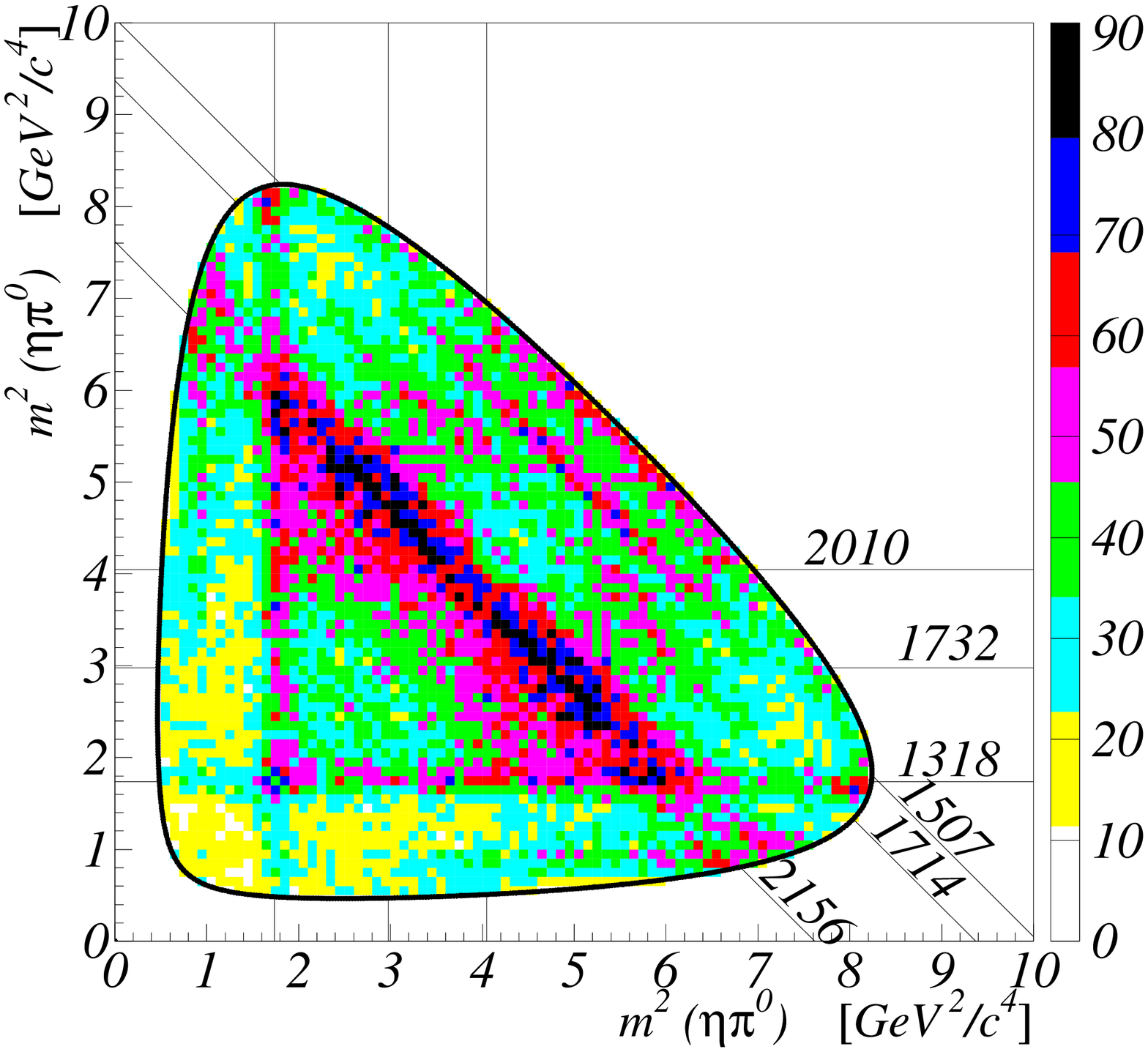}
\put(-1.4,4.2) {\makebox{(a)}}
\put(4.0,4.2) {\makebox{(b)}}
\includegraphics[height=.23\textheight]{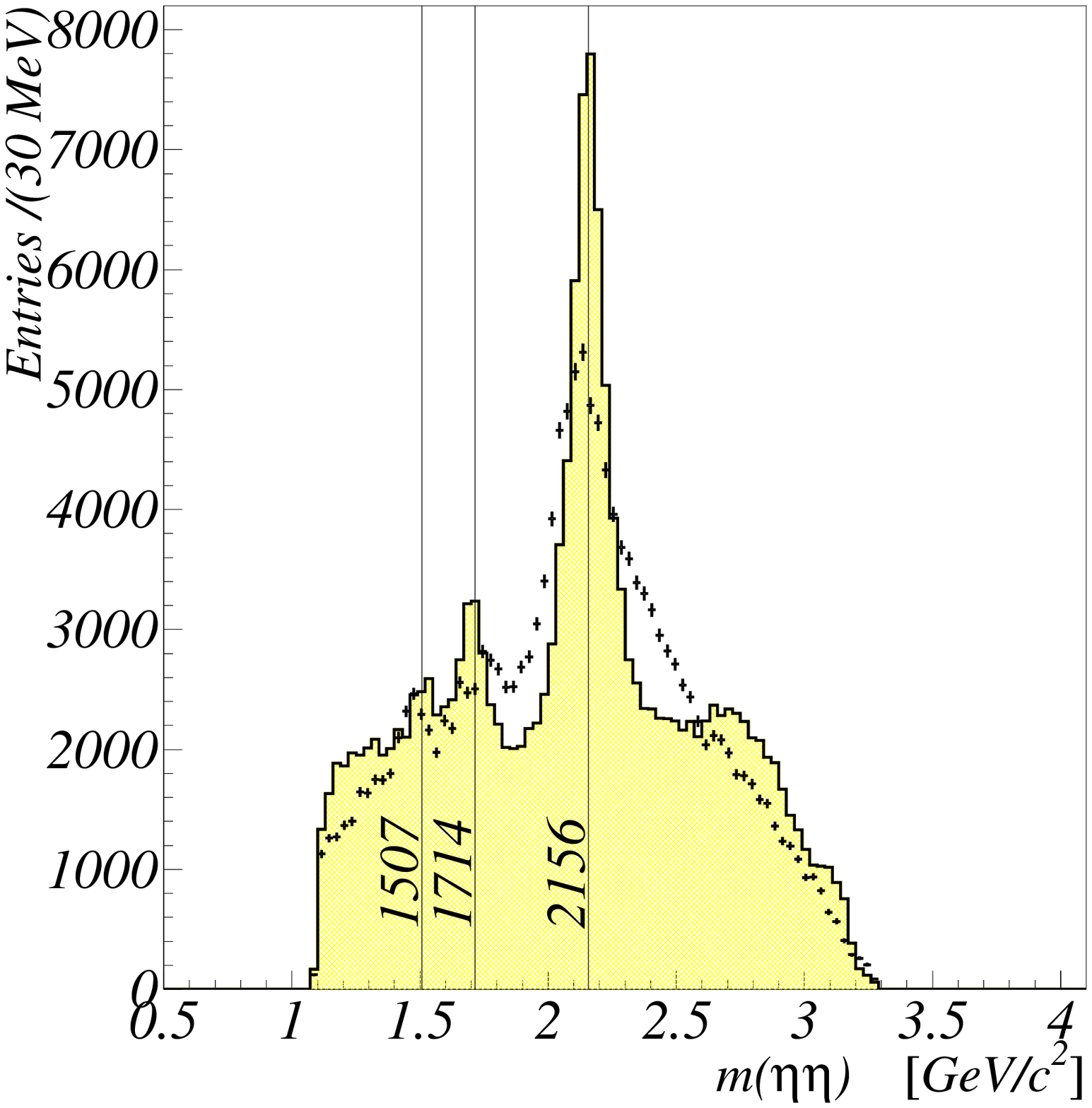}
\put(4.0,4.2) {\makebox{(c)}}
\includegraphics[height=.23\textheight]{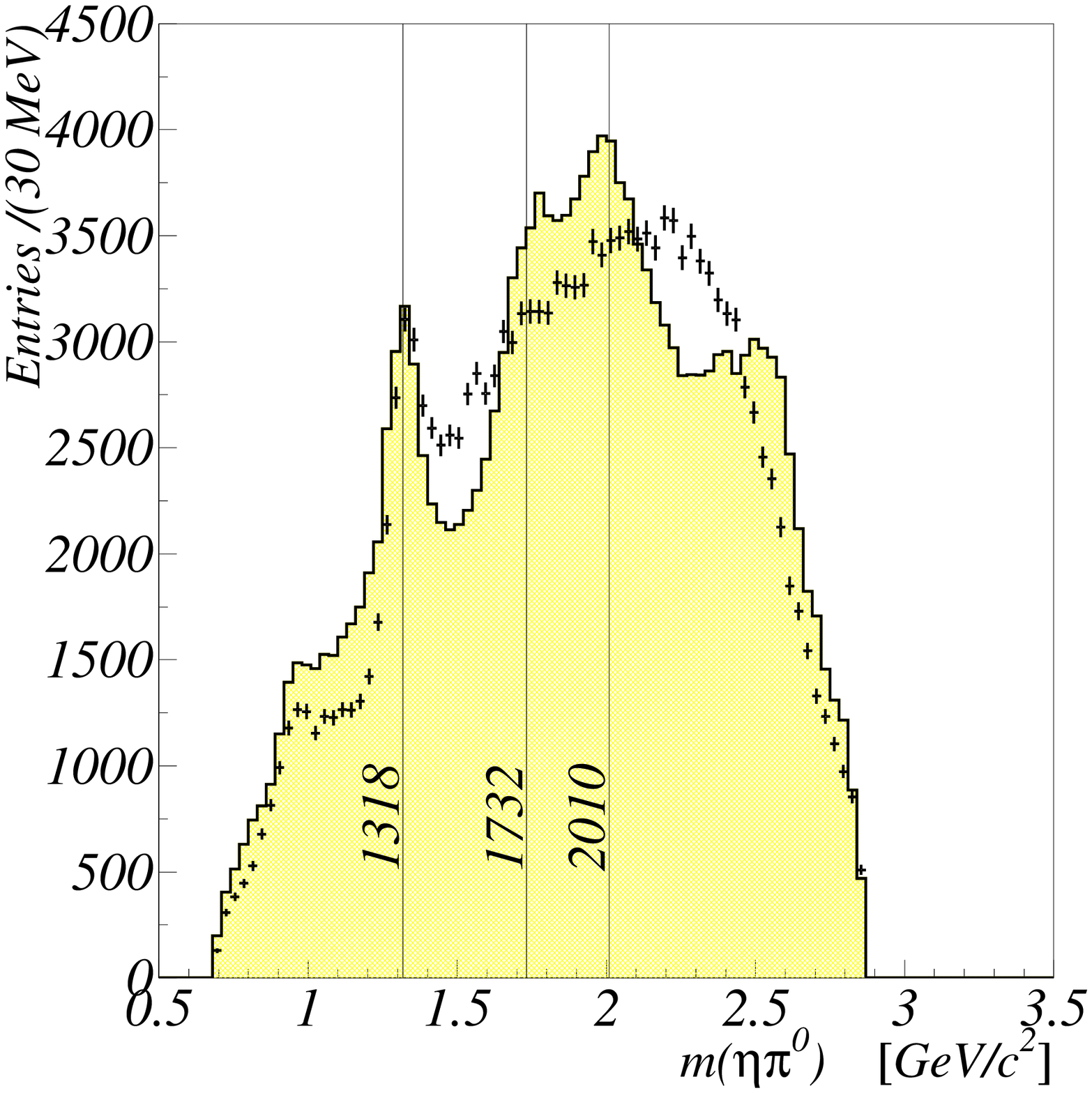}
\caption{(a) Dalitz plot of $\eta\eta\pi^0$. 
(b) and (c) show $\eta\eta$ and  $\eta\pi^0$ mass projections.
The positions of the resonances used in the initial fit are indicated at their PDG mass values in MeV.
The histograms show the relatively poor fit obtained with this initial set of resonances, 
as described in the text.
}
\label{dalitz}
\end{figure*}

The reaction studied was 
\begin{eqnarray}
\bar p p\rightarrow \pi^0\eta\eta, 
\quad \pi^0\rightarrow2\gamma, 
\quad \eta\rightarrow2\gamma
\nonumber
\end{eqnarray}
Twenty two million six photon events were recorded in the full 
detector, of which 3.25 million were fully contained in the 
fiducial volume, had no photons in the forward calorimeter, 
and met the energy-momentum (4C) constraint with probability $>10\%$.
Of these, 139,500 events passed $\eta\eta\pi^0$ seven constraint
fit with $CL>10\%$. These events were then subjected to three
'anticuts'. Events which met the hypotheses for the 'contaminant' 
final states ($\pi^0\pi^0\pi^0$), ($\eta\pi^0\pi^0$) or ($\eta\eta\eta$)
with $CL>2\%$ were removed, and the final sample of 83,400 events 
was used for the Dalitz plot analysis. The Dalitz plot and its 
$\eta\eta$ and $\eta\pi^0$ projections are shown in Fig. \ref{dalitz}.
These mass distributions differ completely from the phase space distributions which are predicted to be essentially flat over the entire $1.0-2.6$ GeV/$c^2$ mass region for both $M(\eta\eta)$ and $M(\eta\pi^0)$.
Visible structures appear to correspond in the $\eta\eta$ 
diagonal to $f_0(1500)$, $f_0(1710)$,  and a complex of states 
in the mass region of $\simeq 2000-2250$ MeV. The visible horizontal 
and vertical $\eta\pi^0$ bands appear to correspond to $a_0(980)$
and a state around 1300 MeV. 
It is important to keep in mind that the 'anticuts' 
described above remove events and introduce distortions
in specific regions of the Dalitz plot.
The effect of the $3\pi^0$ 'anticut' was large, but 
mostly confined to the diagonal edge of the Dalitz plot in the $f_2(1270)$
region.
The effect of the $\pi^0\pi^0\eta$ 'anticut' was smaller 
and also largely confined to the $f_2(1270)$ and $f_0(1370)$ regions. 
The effect of $3\eta$ 'anticut' was small, and was 
mainly in the vicinity of $a_2(1320)$. 

Monte Carlo simulation were done to estimate feed-down contributions 
from 7 and 8 photon reaction with one and two missing photons, respectively, 
which remain in our final event selection. 
It was found that the contributions of  $\omega\eta\pi^0$ ($5.4\pm0.2$) \%,
and  $\omega\pi^0\pi^0$ ($5.2\pm0.2$)\%, 
were mostly confined to the diagonal edge of the Dalitz plot, in the region of 
$f_2(1270)$ and $f_0(1370)$. The contributions of 
$4\pi^0$ was estimated as ($2.3\pm0.1$)\%, and was mostly confined to the 
region around $a_2(1320)$. The overall efficiency of our 
$\eta\eta\pi^0$ selection 
was estimated by MC simulation to be 
($4.26\pm0.1$)\%. 
Taking account  of the feed-down contributions in the total yield,
we obtain the total cross section corresponding to the 
$\eta\eta\pi^0$ events in the Dalitz plot,  
$\sigma(\eta\eta\pi^0)=150\pm1 (stat)\pm2(syst)$ $nb$.

\begin{figure*}[!tb]
\setlength{\unitlength}{1cm}
\includegraphics[height=.23\textheight]{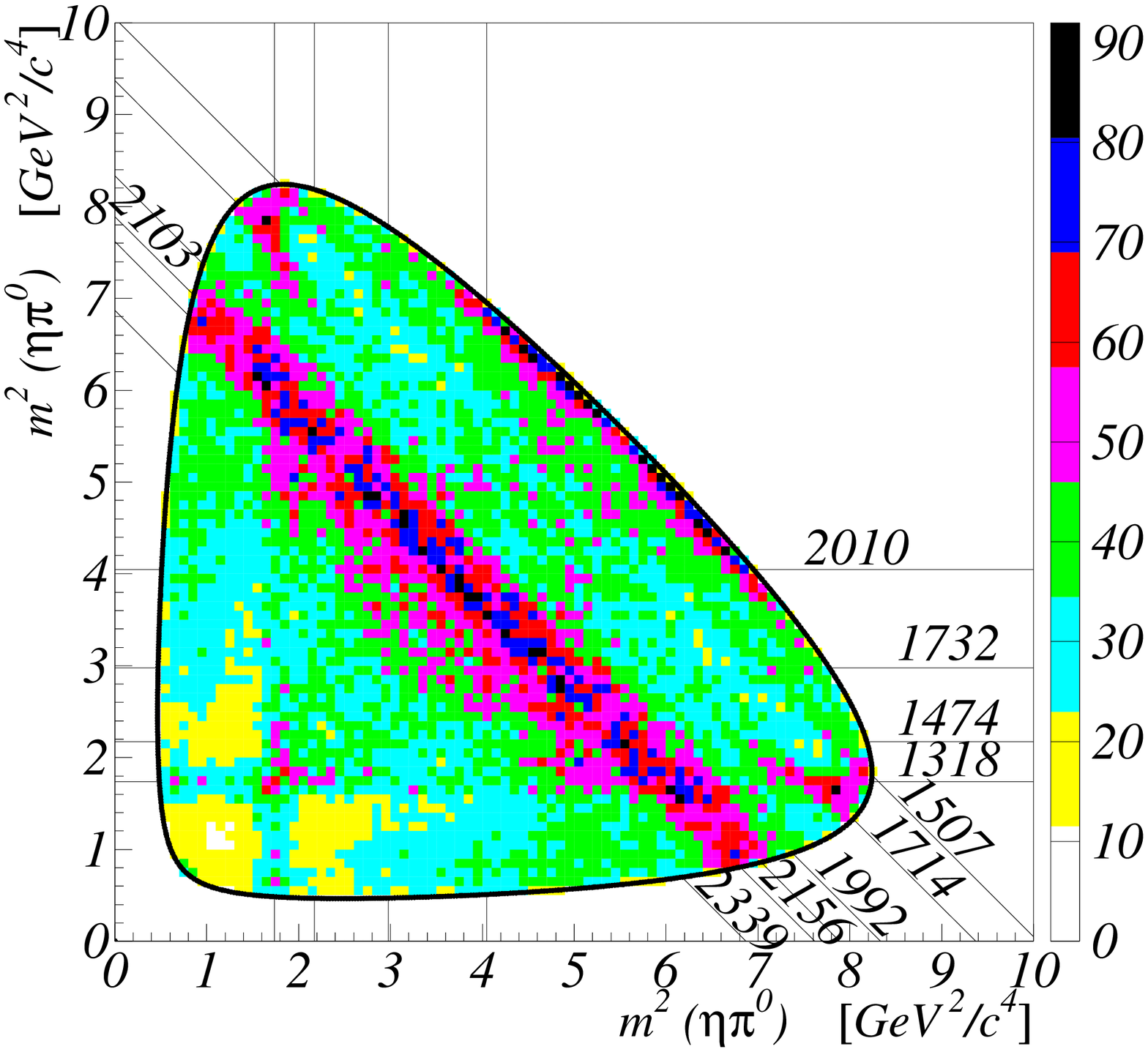}
\put(-1.4,4.2) {\makebox{(a)}}
\put(4.0,4.2) {\makebox{(b)}}
\includegraphics[height=.23\textheight]{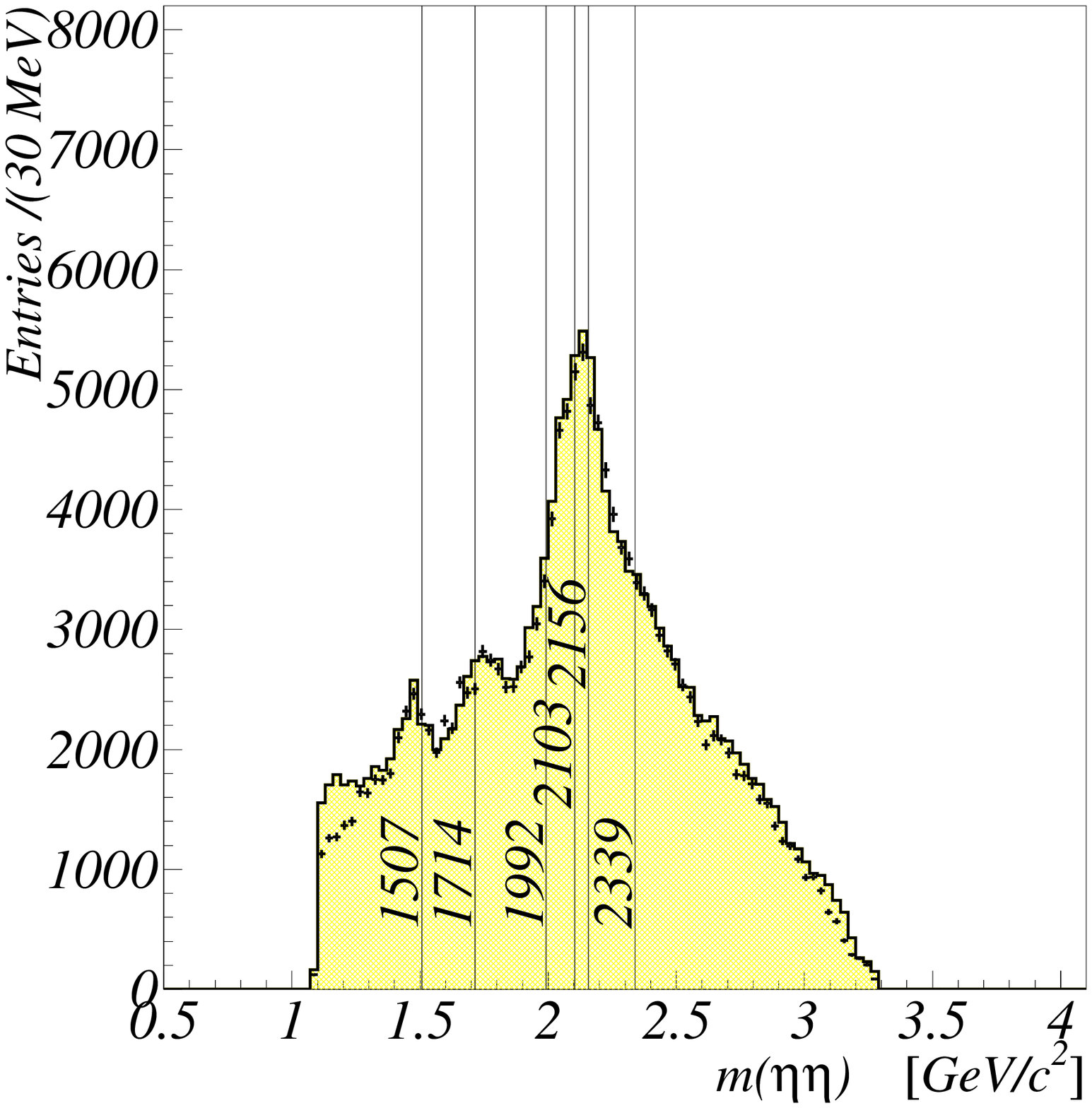}
\put(4.0,4.2) {\makebox{(c)}}
\includegraphics[height=.23\textheight]{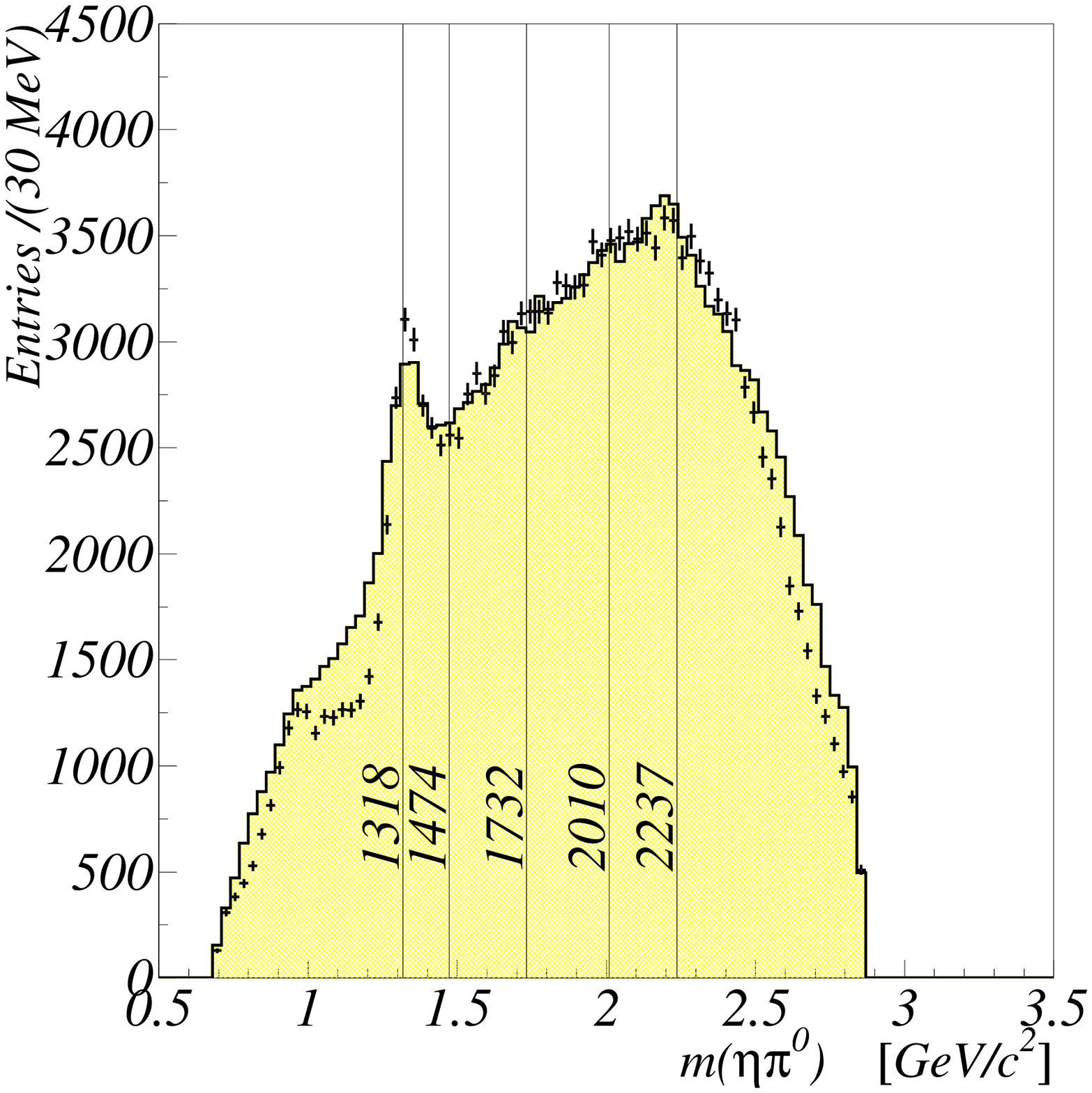}
\vspace*{-5pt}
\caption{(a) The best fit for the Dalitz Plot, 
and the corresponding invariant mass 
projections, (b) $\eta\eta$, and  (c) $\eta\pi^0$. Data are shown with points. 
Histograms show the best fit using the final set of resonances.}
\label{best}
\end{figure*}

\section{Formalism for Dalitz-Plot Analysis}

The reaction  $\bar p p \rightarrow \eta\eta\pi^0$ is described in the 
isobar model as a two step process, $\bar p p \rightarrow f_J+\pi^0$, or   $a_J+\eta$, followed by $f_J \rightarrow \eta\eta$, and $a_J\rightarrow\eta\pi^0$.
In principle both the production and 
the decay of $f_J$ and $a_J$ should be independently taken into 
account. 
In practice, this is not possible.  At an incident antiproton momentum of 5.2 GeV/$c$, $p\bar{p}$ annihilation involves angular momentum up to $\sim7$ in the initial state.  This makes it essentially impossible to use the full helicity formalism for the analysis of our data.  Instead, we use the decay formalism developed and successfully used in several Dalitz plot analyses of the Crystal Barrel $p\bar{p}$ annihilation data at 1.94 GeV/$c$ \cite{Abele:pf,Adomeit:1996nr,Anisovich:1999}.  In addition, we take account of the production amplitudes in an average way by parameterizing them as a function of even powers of $\cos\theta$, where $\theta$ is the resonance production angle with respect to the $\bar{p}$ direction, $p(\theta)=1+\sum^n_{i=1} p_i \cos^{2i}\theta$.  The final amplitude for the formation and decay of a resonance with spin $J$, projection $\lambda\equiv J_x$, mass $m_0$ and width $\Gamma_0$ is the given by
\begin{eqnarray}
\small
A_J^{\lambda} = p(\theta) G_\lambda  e^{i\delta_\lambda }
F_J(q)
\frac
{Y_J^\lambda (\alpha,\beta)}
{m_0^2-s-im_0\Gamma_m},
\label{ampl}
\end{eqnarray}

\begin{table*}[!tb]
\newcommand{\m}{\hphantom{$*-$}}
\newcommand{\cc}[1]{\multicolumn{1}{c}{#1}}
\renewcommand{\arraystretch}{1.2} 
\begin{tabular}{@{}c|c|c|c|c|c|c|c|c}

\hline
Resonance  &  $\Delta ln{L}/\Delta \# param.$ && 
Interference & $\Delta ln{L}/\Delta \# param.$ &
$c_0$ & $\Delta(\delta_0)$ &  $c_1$ & $\Delta(\delta_1)$ \\
\hline

$f_0(2020)$ & $1464/1$ && 
$a_4(2040)\times a_4(2040)$ & $363/2$  &
-0.51 & 0 &  -1.00 & 0 \\  
\hline

$f_0(2100)$ & $158/1$ && 
$a_4(2040)\times a_2(1320)$ & $168/4$  &
0.83 & 0.06 &  1.00 & 0.41 \\  
\hline

$f_2(2340)$ & $878/2$ && 
$a_4(2040)\times f_2(2150)$ & $155/4$  &
-0.48 & 1.29 & -1.00 & 1.36 \\  
\hline

$a_0(1450)$ & $297/1$ && 
$a_4(2240)\times a_4(2240)$ & $254/2$  &
-0.80 & 0 &  -0.83 & 0 \\  
\hline

$a_4(2240)$ & $267/2$ && 
$a_2(1320)\times f_0(2100)$ & $161/2$  &
-1.00 & 0.17 &  - & - \\    
\hline
\end{tabular}
\caption{Resonances and interferences added in the second iteration.
$c_\lambda$ and $\Delta(\delta_\lambda)$ ($\lambda=0,1$) refer to the interference coefficients and the phase differences (in radians) between the interfering resonances.
}
\label{tab:a}
\end{table*}

where $G_\lambda$ and the $\delta_\lambda$ are the magnitude and the 
phase of the complex coupling constant,
$Y_J^{\lambda}(\alpha,\beta)$ are spherical 
harmonics which are functions of angles $\alpha$ and $\beta$,
the polar and azimuthal angles of the decay in the rest frame 
of the resonance. The denominator is that for the 
relativistic Breit Wigner with energy dependent width,
\begin{eqnarray}
\small
\Gamma_m = \Gamma_0 \left(\frac{q F^2_J(q)}{m}\right)/\left(\frac{q_0 F^2_J(q_0)}{m_0}\right) \quad m=\sqrt s.
\label{blatt}
\end{eqnarray}
Here $F_J(q)$ are the Blatt-Weisskopf barrier factors
for the resonance break-up momentum q (subscripts $0$ denote values at 
the resonance mass, $m_0$), evaluated for a radius of 1 fermi.
The combined intensities of two interfering resonances 1, and 2 in the 
Dalitz plot are given by 
\begin{eqnarray}
 w_{1,2}= & \sum_{\lambda}[|A_{\lambda,1}|^2
+|A_{\lambda,2}|^2 + 2c_\lambda \Re(A_{\lambda,1} A_{\lambda,2}^{*})], 
\label{int}
\end{eqnarray}
where the interference coefficient $c_\lambda$ ranges from 0 
(no coherence) to $\pm 1$ (maximum coherence).
We wish to note here that when two resonances interfere the difference between their phase angles, $\Delta(\delta_A)$ also becomes important.
In considering interferences a certain pragmatic compromise  has 
to be made. With 11 different resonances in our final result 
(see Table \ref{tab:b}),
if all possible two state interferences, including 
self-interference were taken into account, one would need
($60\times\lambda$ multiple) values of $c_\lambda$. This is impossible 
and also unnecessary. In the present work a large number 
of possible interferences were studied, but only those were retained in the 
final analysis which improved the log-likelihood by more than 20 per additional   
degree of freedom. The negative log-likelihood of a MC fit to the data was defined as 
\begin{eqnarray}
-ln{L} =Nln(\sum_{i=1}^M w_i^{MC})- (\sum_{i=1}^N ln w_i^{DATA}), 
\label{nllikeli}
\end{eqnarray}
where $w_i^{DATA}$ refer to the N events in the data, and 
$w_i^{MC}$ refer to the M events in the MC fit.

With this definition an increase of $\Delta (ln{L})=(0.5\times r)$
with increase of r degrees of freedom corresponds to a one 
standard deviation ($\Delta\chi^2/d.o.f.=1$) improvement 
in the fit.
For making the final fits $m_0$, $\Gamma_0$ and $J$ are manually varied,
$p(\theta)$ are separately fit, and the free parameters for the fit are 
$G_\lambda$, $\delta_\lambda$ and $c_\lambda$. 
As detailed below, in practice much more demanding criteria for $\Delta (\ln L)$ were used.

The ingredients of fitting the Dalitz plot are masses, widths, 
$J^{PC}$, the production amplitude parameters
$p(\theta)$, and resonance-resonance interferences including 
self-interference. We take these into account successively. We start with 
a basic set of six resonances. These include those which are identifiable 
as clear enhancements in the Dalitz plot, $f_0(1500)$, $f_0(1710)$, and
$f_2(2150)$ as representative of the broad enhancements in the  
$M(\eta\eta)\sim2000-2200$ MeV region, and $a_2(1320)$. In this basic 
set we also include $a_2(1700)$ and $a_4(2040)$ which are not apparent
as identifiable enhancements, but which are considered well established 
and whose decay in $\eta\pi^0$ has been observed \cite{Eidelman:2004wy}.

The initial fit with this set of resonances was done at the simplest level.
Their masses and widths were fixed at PDG values \cite{Eidelman:2004wy},
neither the the angular distribution of the production 
amplitudes, nor resonance-resonance interference was included, and only 
the ten $G_\lambda$ were fitted by maximizing $-ln{L}$.
The fit was quite poor, with $ln{L}/ \#par.=-9,520/10$.

As a first step in improving the fit, it was decided to take account of the fact that at 5.2 GeV/c the production amplitudes are far from 
isotropic. 
It was found that their angular distributions could be satisfactorily 
parameterized  as a third order polynomial in $cos^2(\theta)$. 
This added 18 parameters in the fit which improved by  
$\Delta ln{L}/ \Delta \#par.=5,261/18$.
Inclusion of the strong self-interference of $a_4(2040)$  led to an 
additional $\Delta ln{L}/ \Delta \#par.=363/2$, bringing the net 
$-ln{L}/ \#par.=15,144/30$. The resulting fit is shown in Fig. 
\ref{dalitz}-(b) and (c), for the projections of invariant masses $M(\eta\eta)$ and 
$M(\eta\pi^0)$, respectively. 
Both are rather poorly fit, and it is obvious that 
additional resonances and interferences must be considered.

The additional resonances input include $f_0(2020)$, $f_0(2100)$,
$f_2(2340)$, and $a_0(1450)$, which are not considered by PDG as 
firmly established, 
and a new state $a_4(2240)$. As shown in Table \ref{tab:a},
the log-likelihood improvements due to the inclusion of these resonances 
is in all cases more than $130$ per added parameter.
Further, only those interferences are included which increase 
log-likelihood by more than $\sim 40$ per added parameter.
Fig. \ref{best} shows the much improved fit to the mass projections.

For the finally chosen resonances two dimensional optimization 
of the mass and width was done for different $J^{PC}$ assignments.
The final result of the optimization was
$-ln{L}/ \#par.=19894/44$.

\vspace*{20pt}

\begin{figure*}[!tb]
\setlength{\unitlength}{1cm}
\includegraphics[height=.23\textheight]{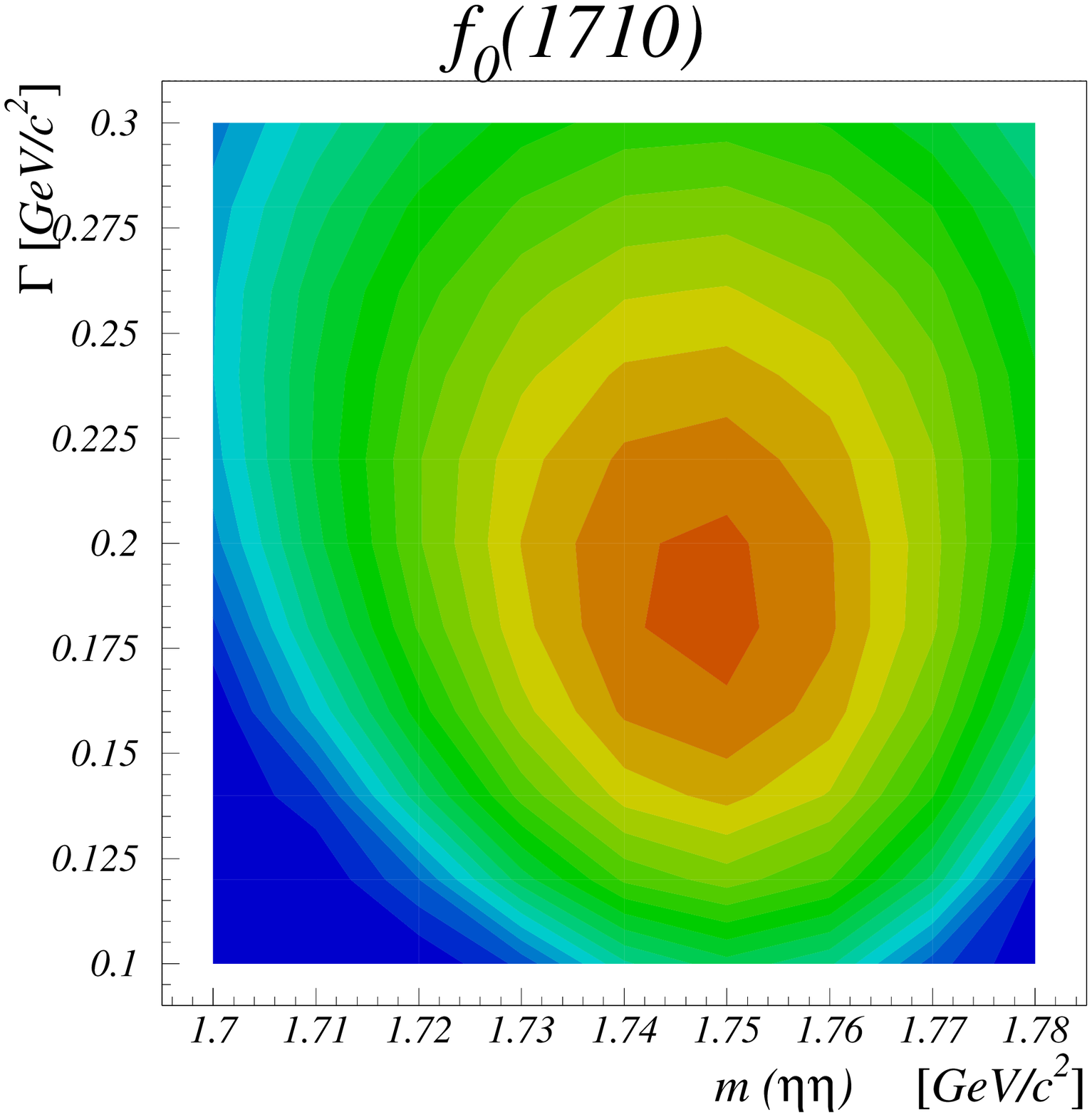}
\put(-4.3,3.7) {\makebox{(a)}}
\put(1.1,4.4) {\makebox{(b)}}
\includegraphics[height=.23\textheight]{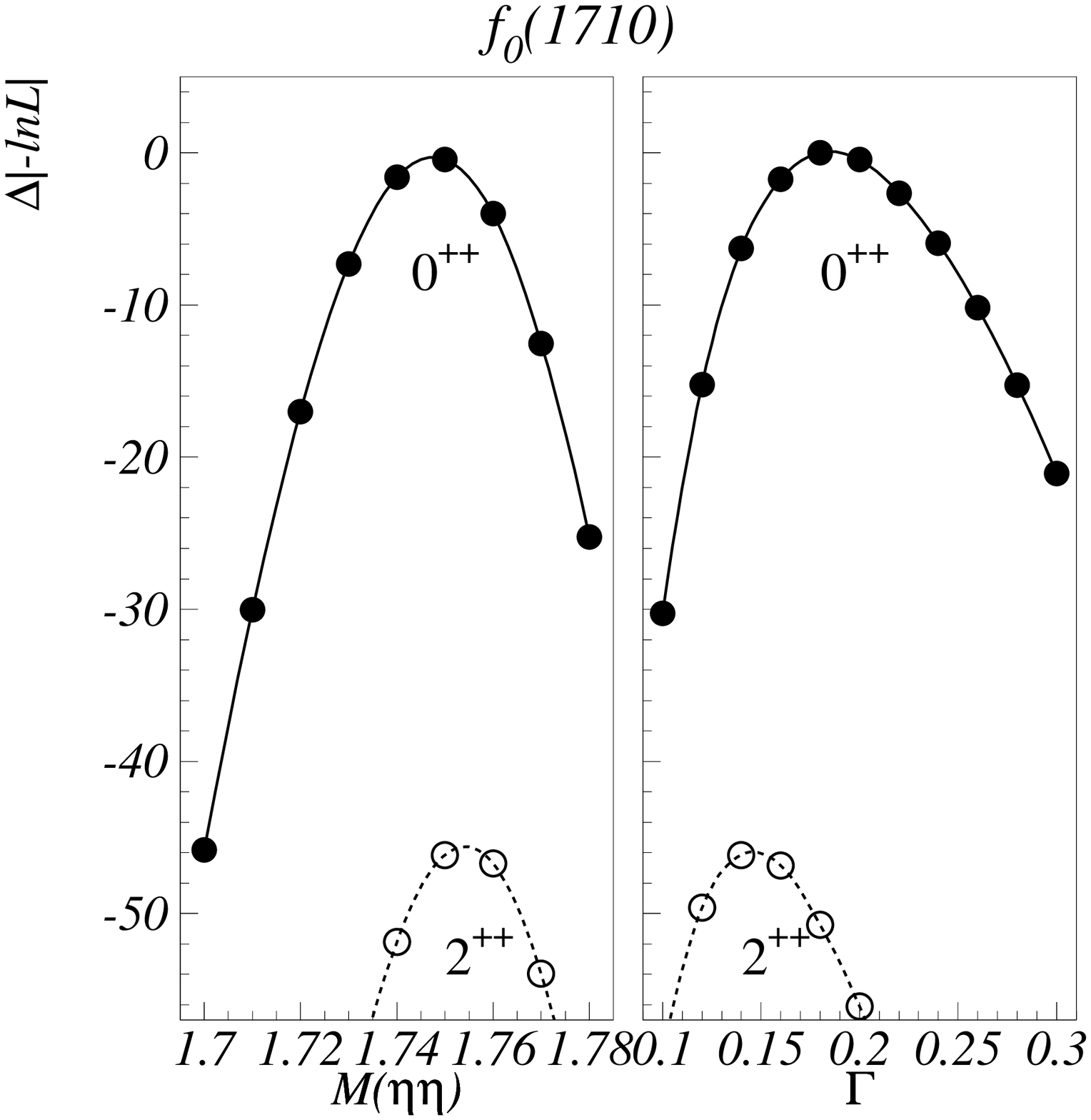}
\put(1.1,4.4) {\makebox{(c)}}
\includegraphics[height=.23\textheight]{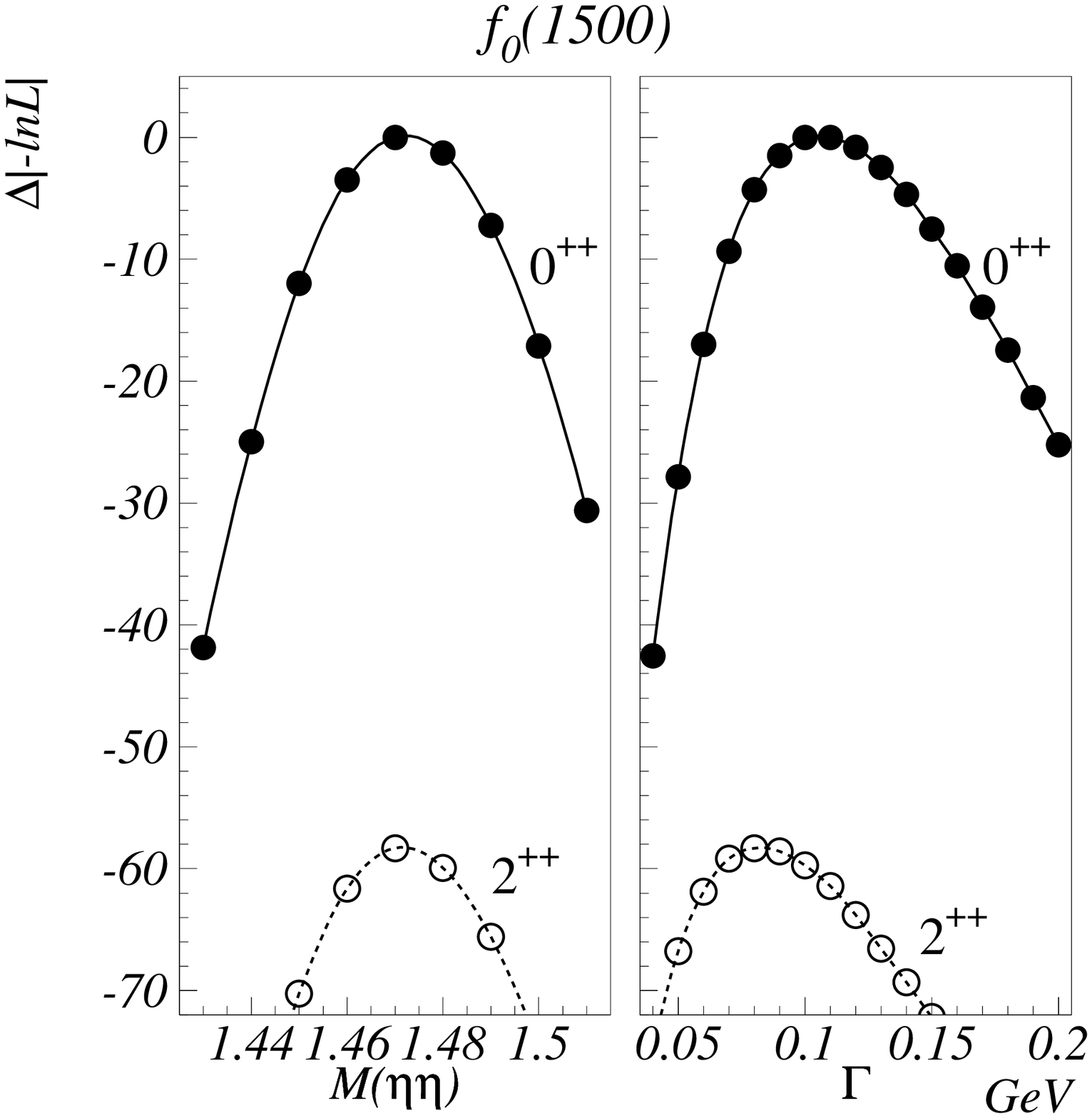}
\setlength{\unitlength}{1cm}
\vspace*{-5pt}
\caption{Results for $\Delta |-ln{L}|$ optimization of mass and width: 
(a) contour plot for $f_0(1710)$, 
(b) $\Delta |-ln{L}|$ variation for mass and width 
for $f_J(1710)$, $J^{PC}=0^{++}$  and 
$2^{++}$, (c) same for $f_J(1500)$}
\label{f01710}
\end{figure*}
\begin{figure*}[!tb]
\setlength{\unitlength}{1cm}
\includegraphics[height=.23\textheight]{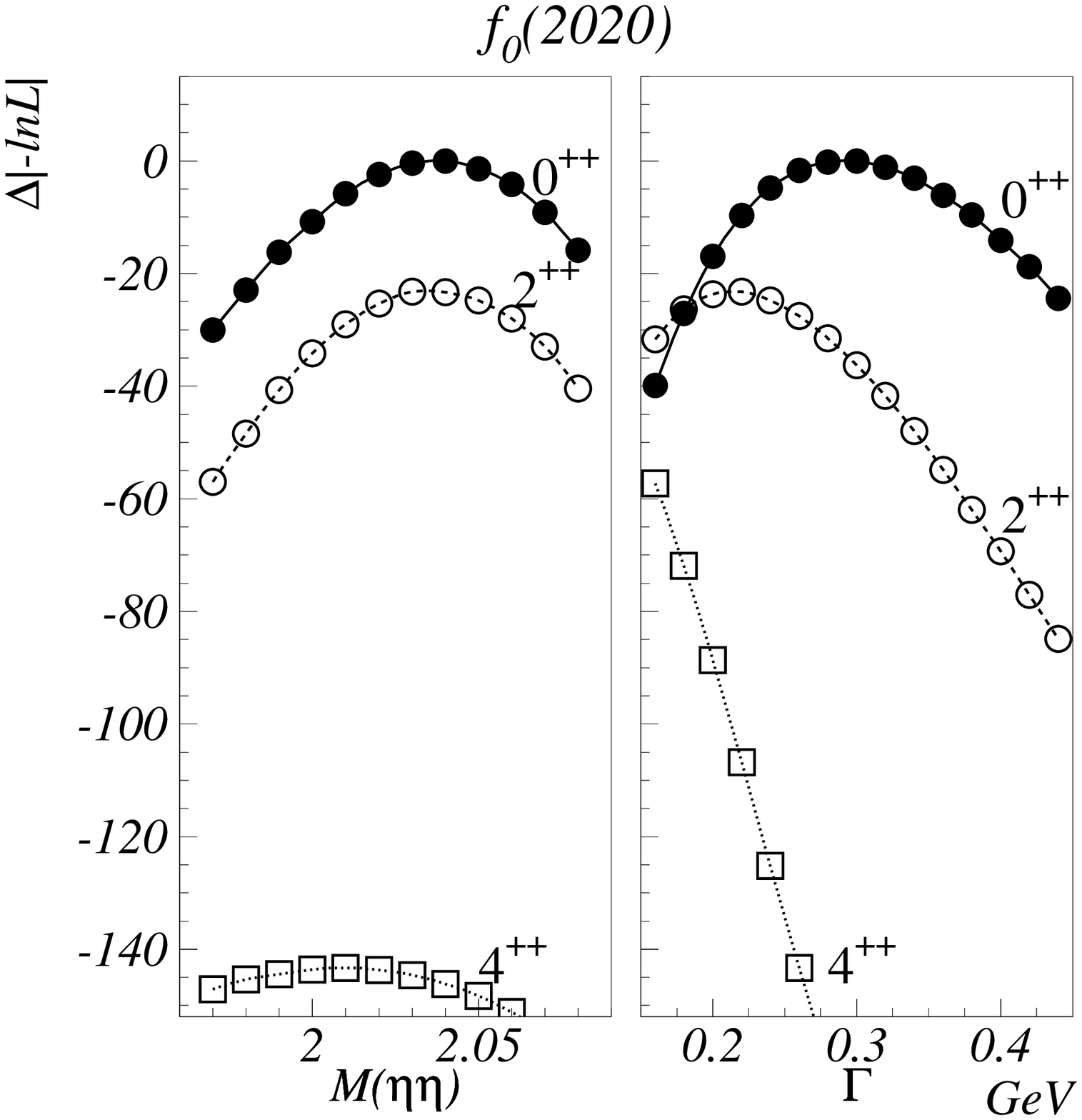}
\put(-4.3,4.4) {\makebox{(a)}}
\put(1.1,4.4) {\makebox{(b)}}
\includegraphics[height=.23\textheight]{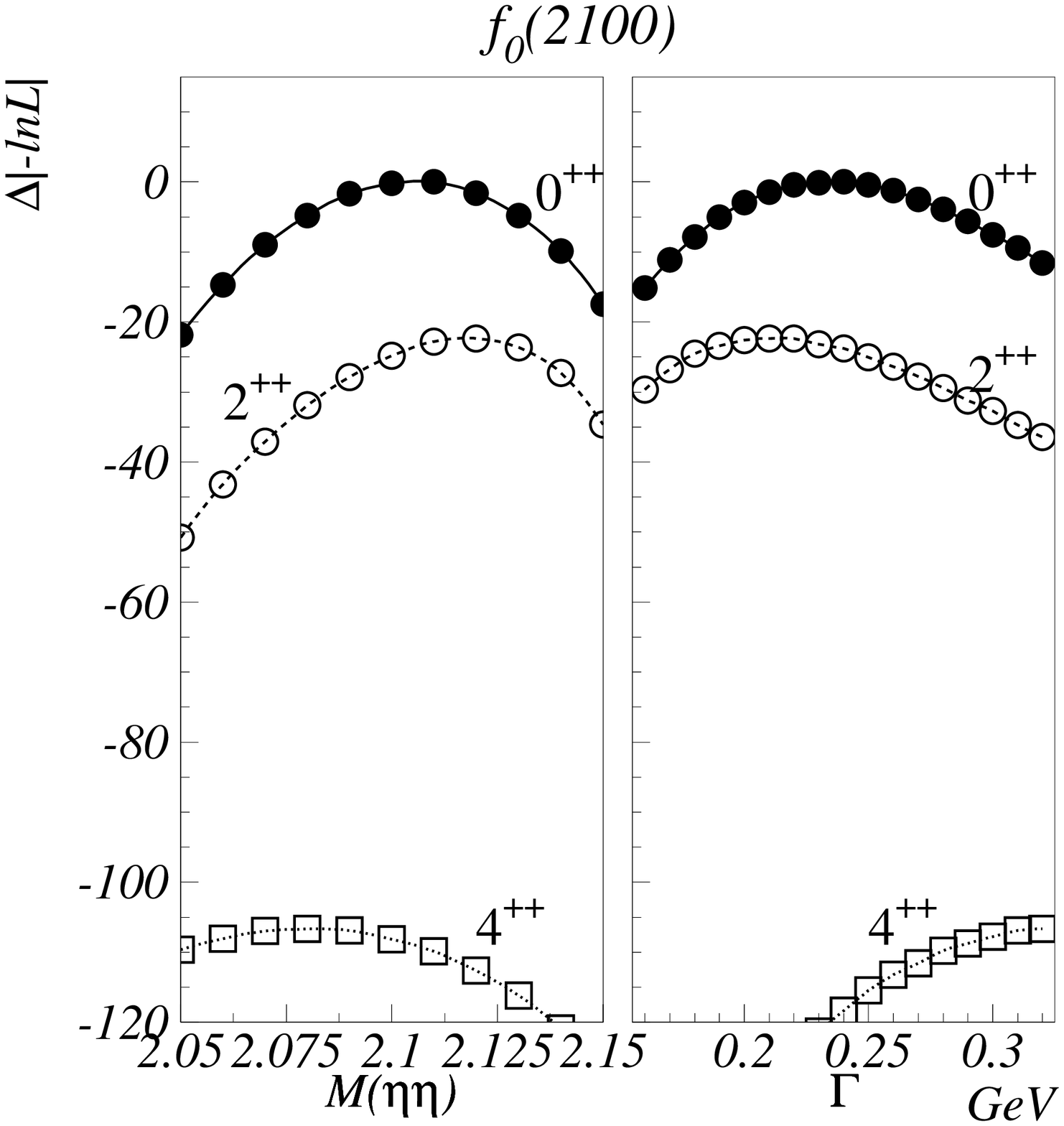}
\put(1.1,4.4) {\makebox{(c)}}
\includegraphics[height=.23\textheight]{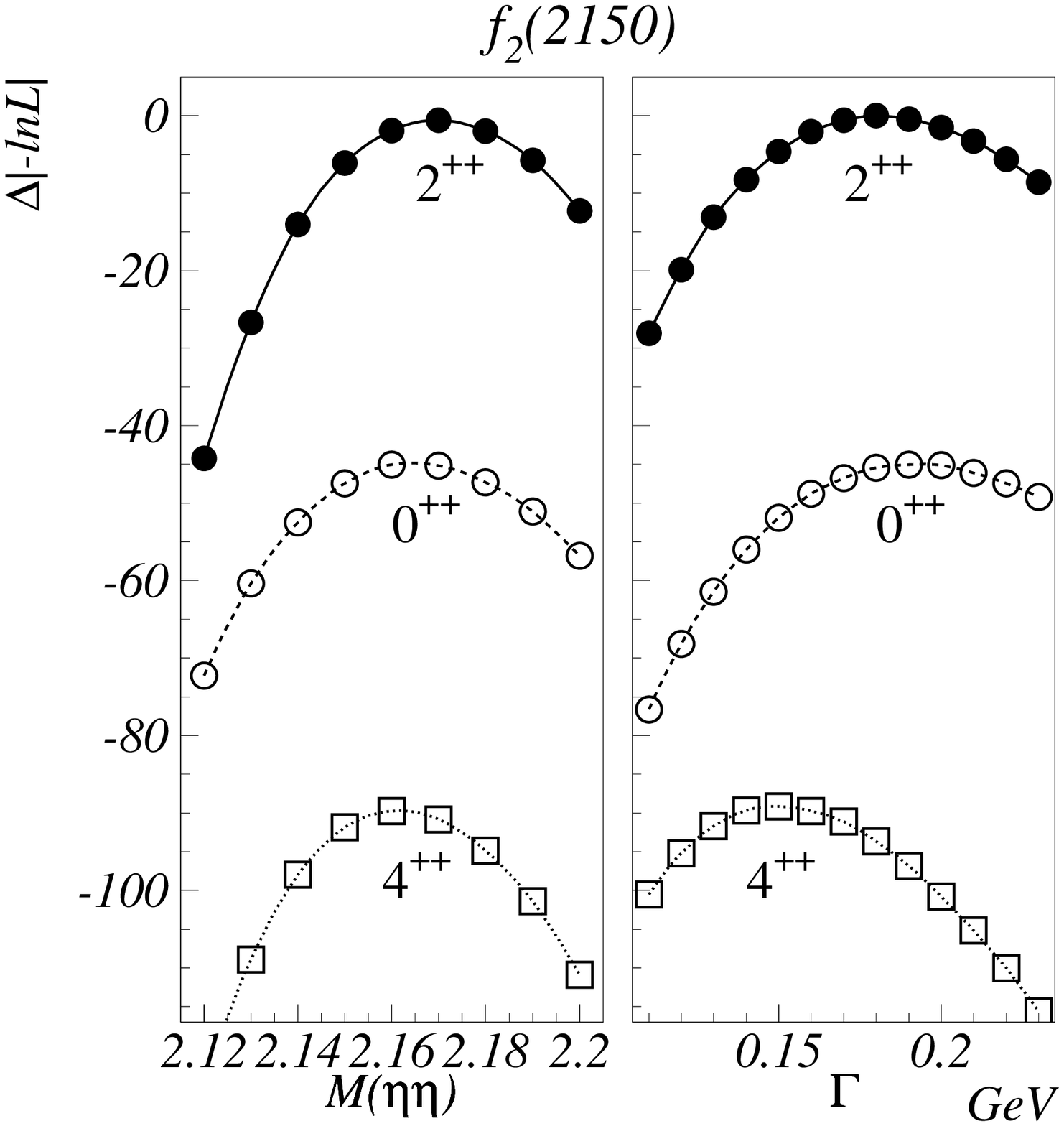}
\vspace*{-5pt}
\caption{Results for $\Delta |-ln{L}|$ optimization for 
(a) $f_0(2020)$, (b) $f_0(2100)$, and $f_2(2150)$.
}
\label{f02020f22150}
\end{figure*}
\begin{figure*}[!tb]
\setlength{\unitlength}{1cm}
\includegraphics[height=.23\textheight]{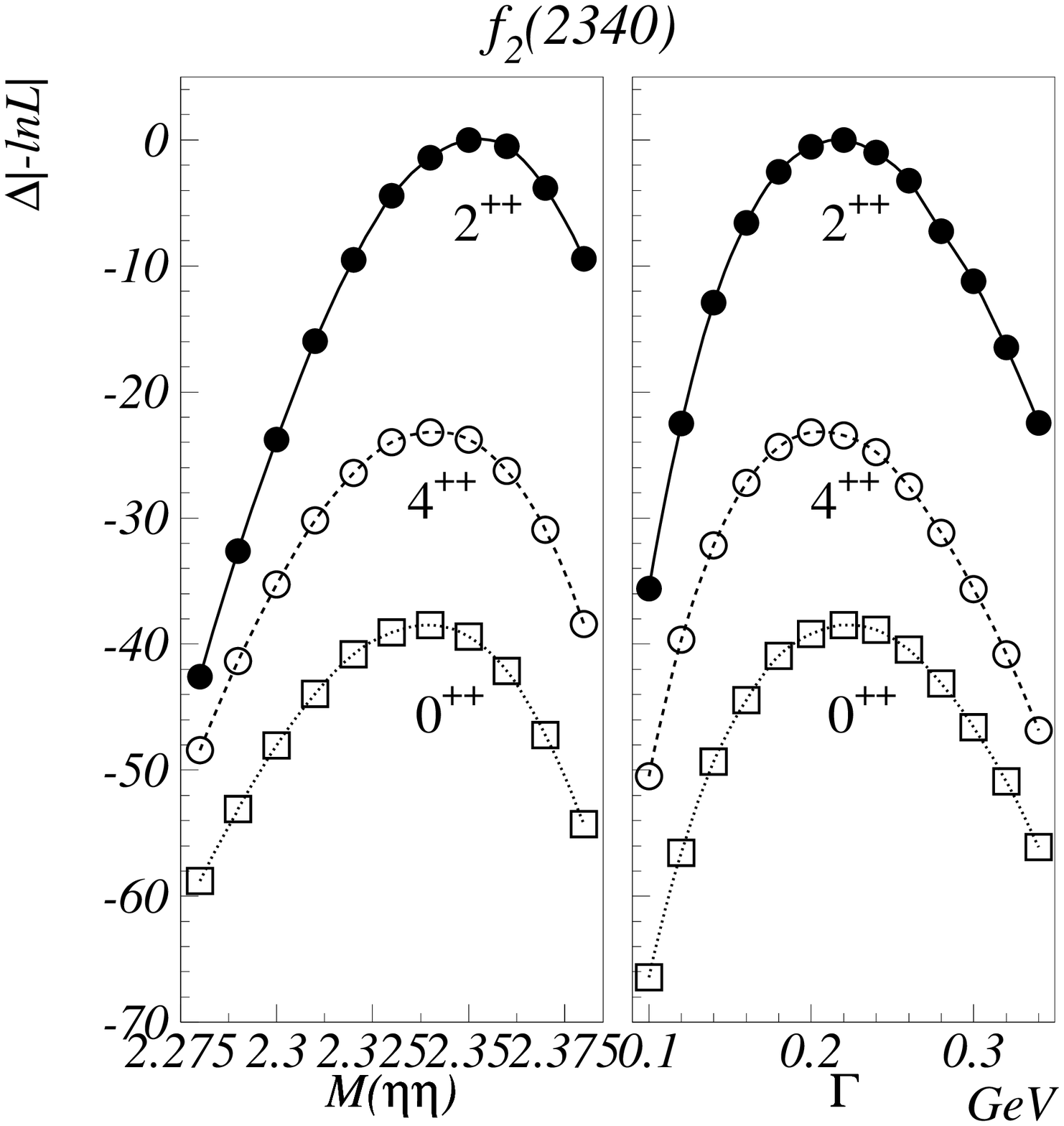}
\put(-4.3,4.4) {\makebox{(a)}}
\put(1.1,4.4) {\makebox{(b)}}
\includegraphics[height=.23\textheight]{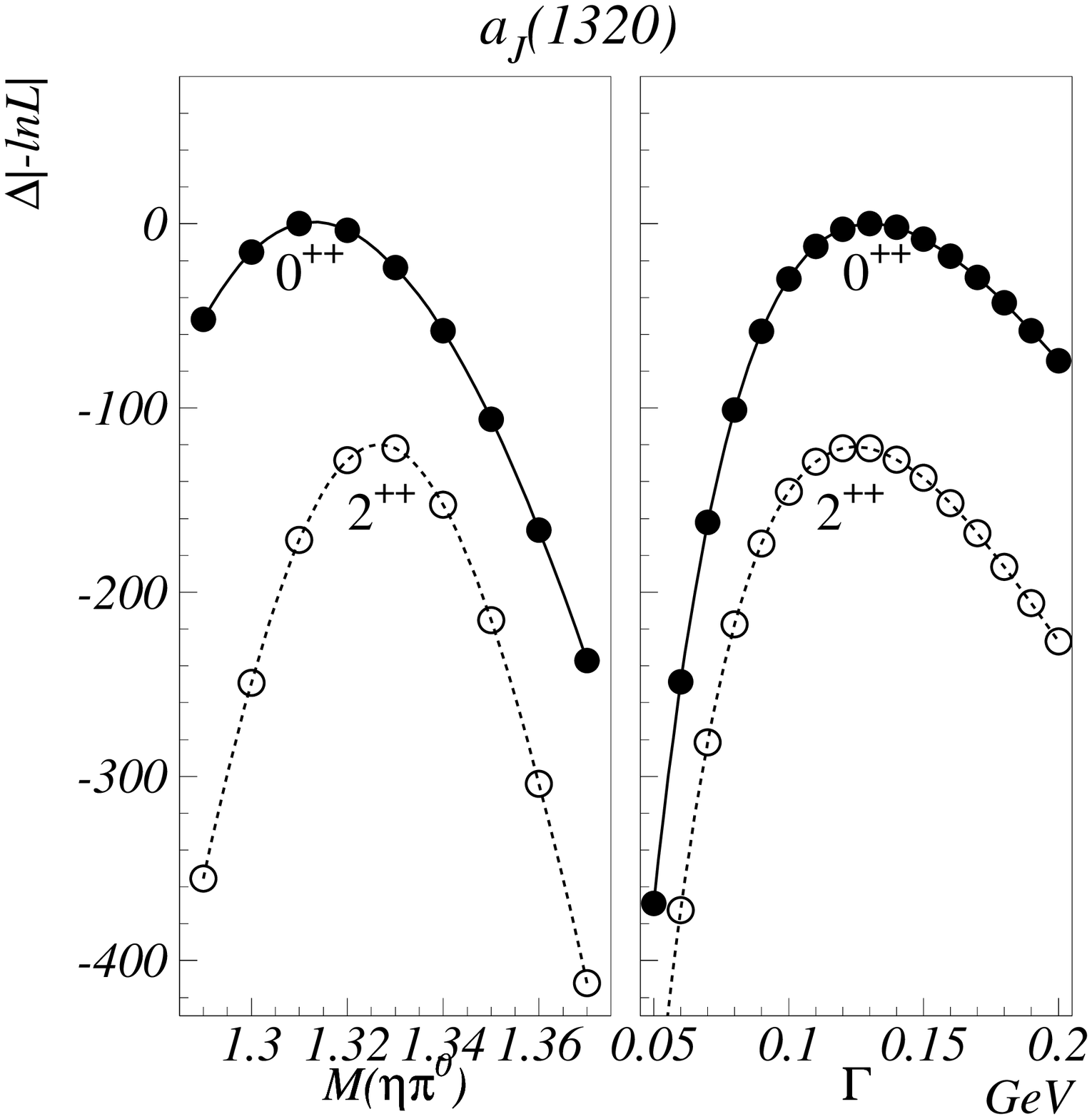}
\put(1.1,4.4) {\makebox{(c)}}
\includegraphics[height=.23\textheight]{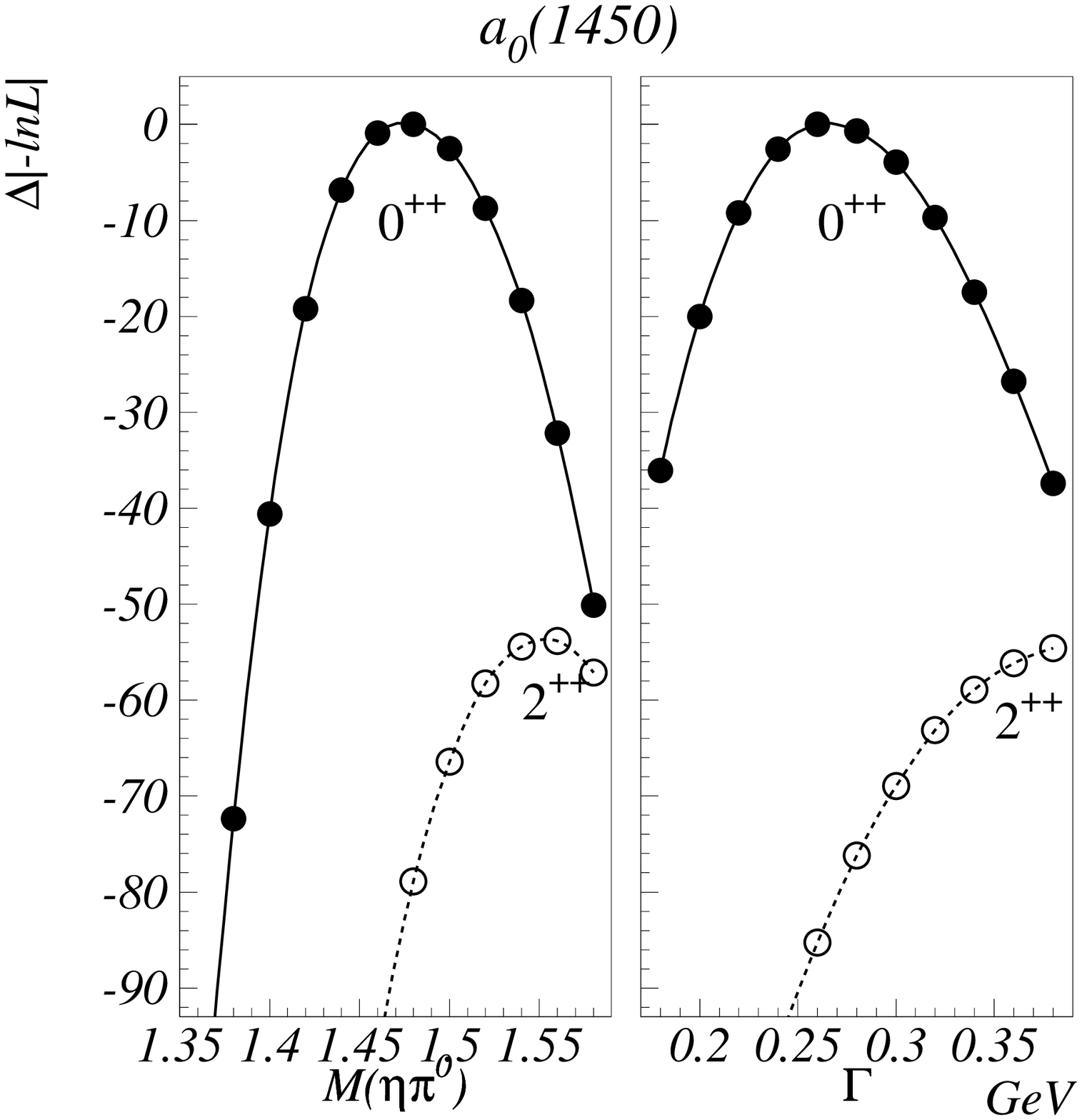}
\vspace*{-5pt}
\caption{ Results for $\Delta |-ln{L}|$ optimization for 
(a) $f_2(2340)$, (b) $a_J(1320)$, and $a_0(1450)$.
}
\label{f22300a01450}
\end{figure*}
\begin{figure*}[!tb]
\setlength{\unitlength}{1cm}
\includegraphics[height=.23\textheight]{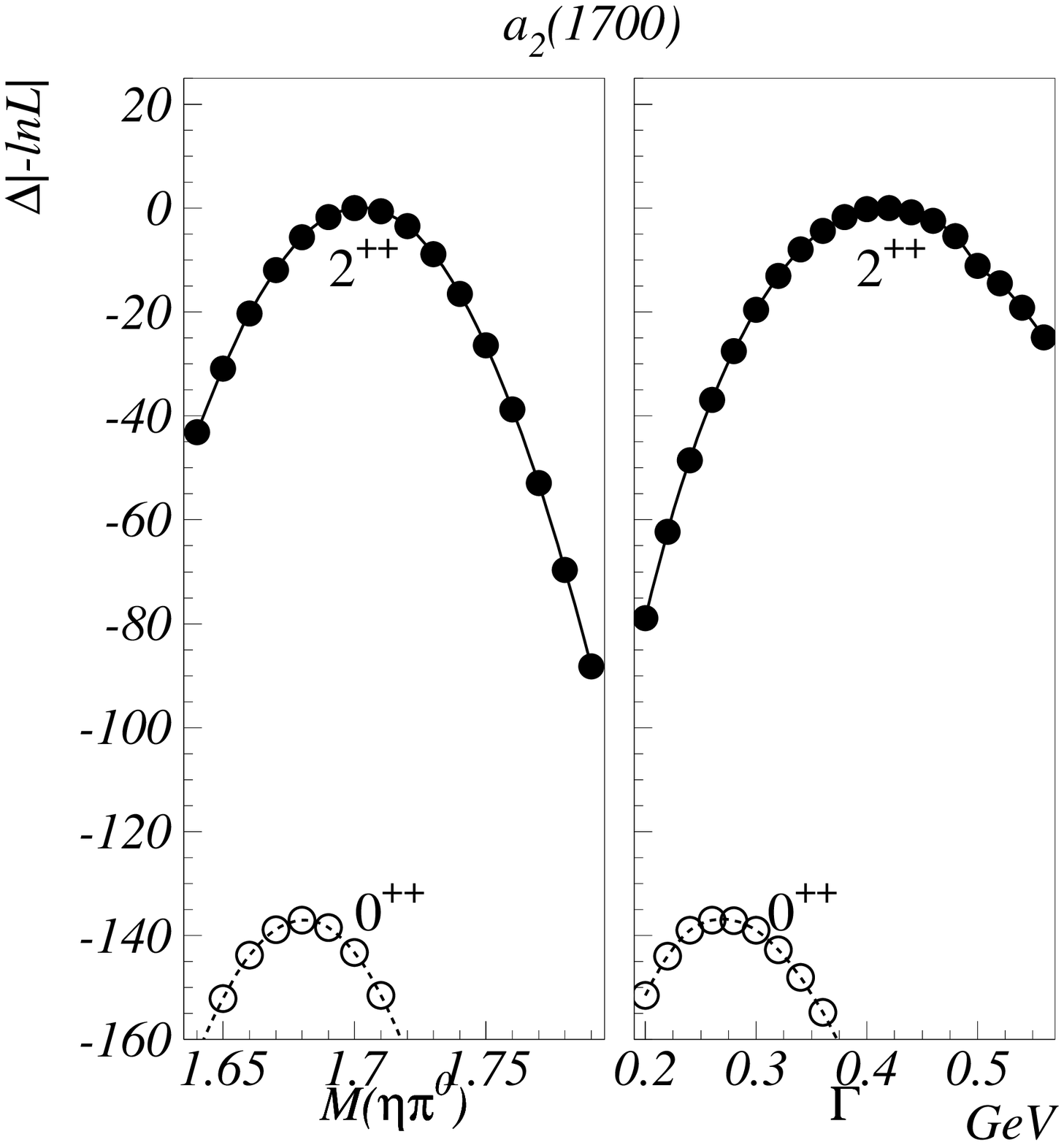}
\put(-3.8,4.4) {\makebox{(a)}}
\put(1.6,4.4) {\makebox{(b)}}
\includegraphics[height=.23\textheight]{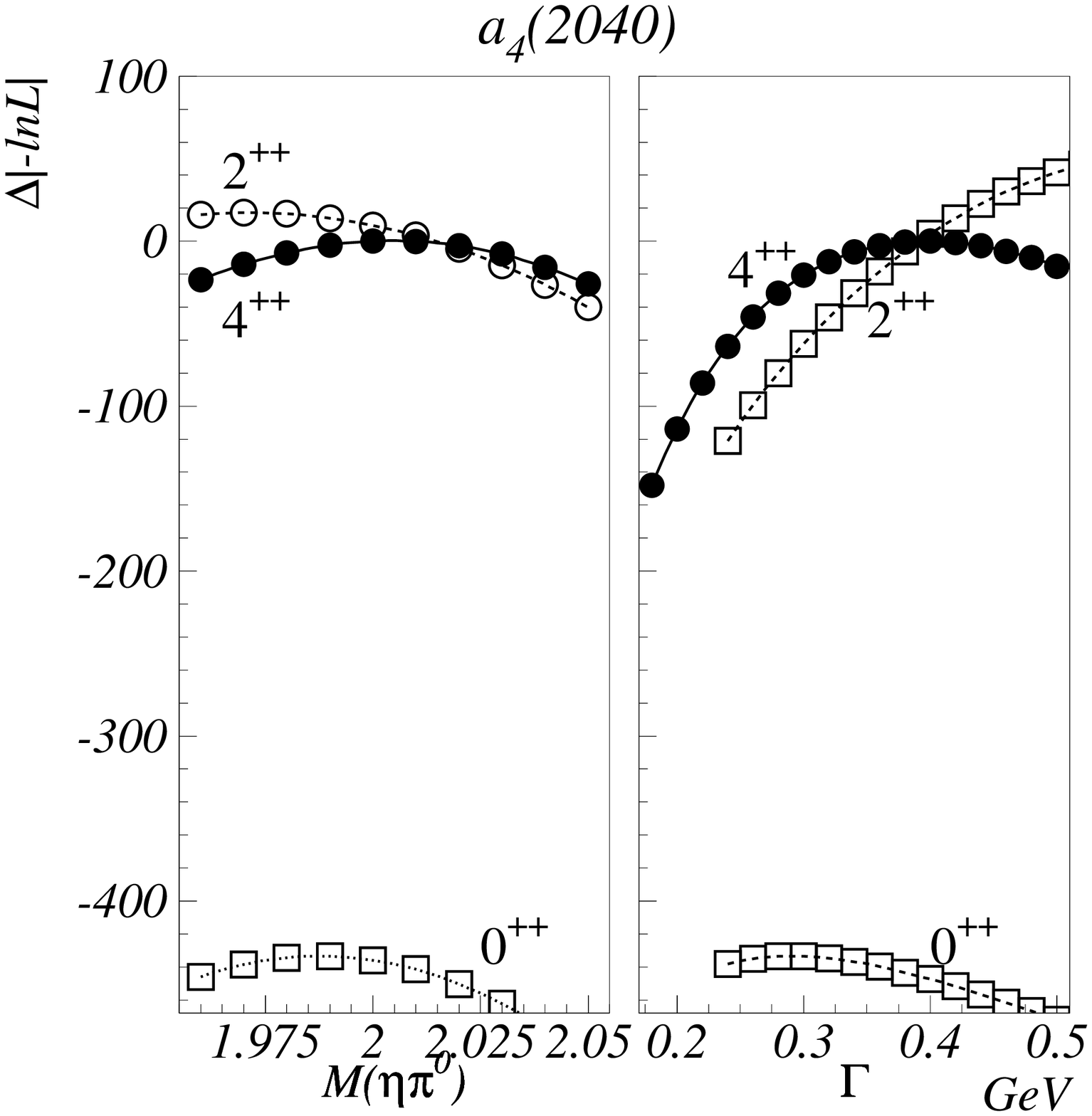}
\put(1.1,4.4) {\makebox{(c)}}
\includegraphics[height=.23\textheight]{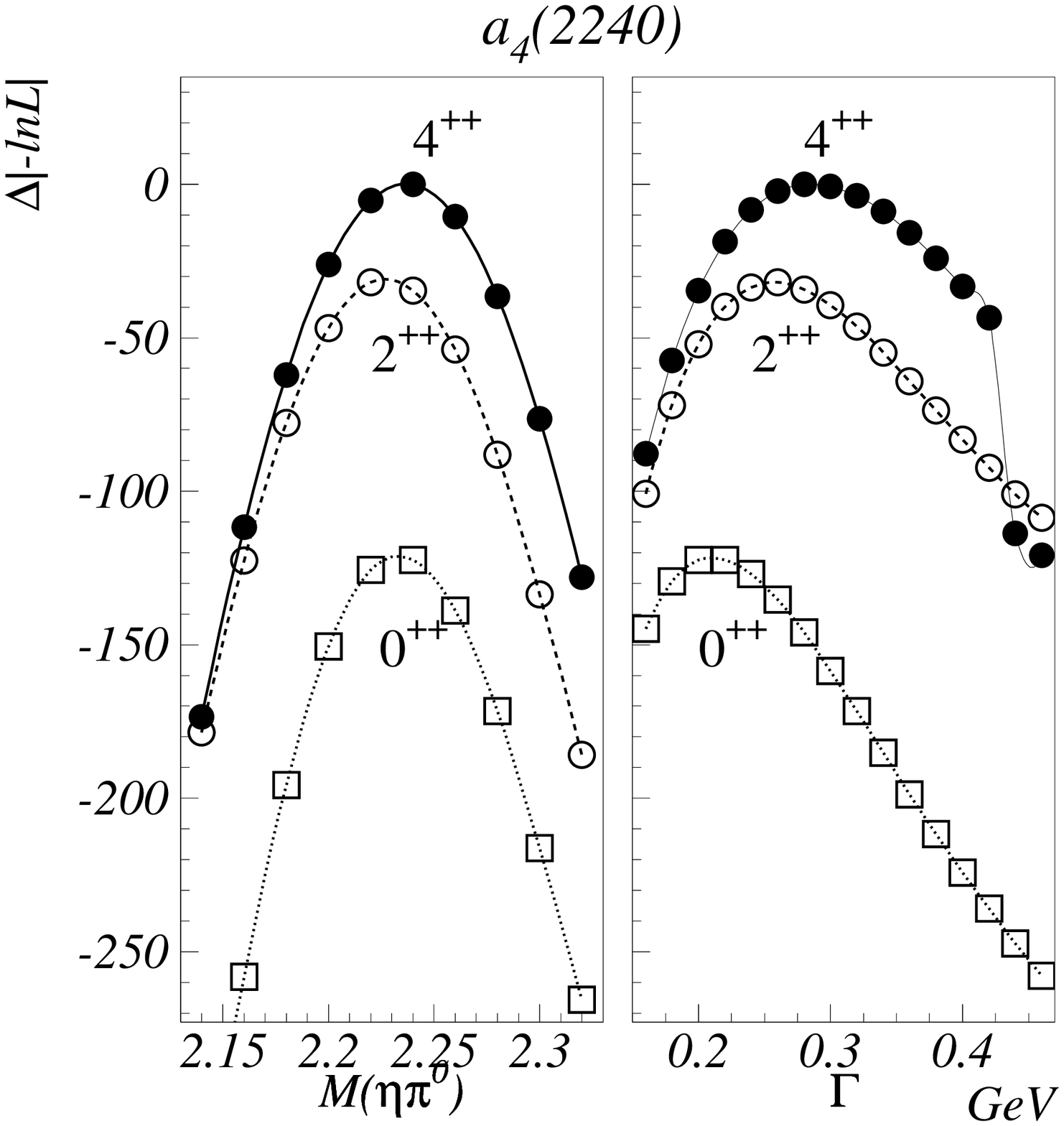}
\vspace*{-5pt}
\caption{
Results for $\Delta |-ln{L}|$ optimization for 
(a) $a_2(1700)$, (b) $a_4(2040)$, and $a_4(2240)$.
}
\label{a21700a42240}
\end{figure*}

\section{Results and Discussion}

Fig. \ref{best} shows the overall results of the best fit to the Dalitz 
plot and the mass projections for $M(\eta\eta)$ and $M(\eta\pi^0)$.
Our final results are summarized in Table II, along with the corresponding results from PDG \cite{Eidelman:2004wy}. We wish to note that the relative intensity fractions listed in Table II are related to, but not the same as relative branching fractions.  This is partly due to the approximations inherent in the formalism used in our partial wave analysis, and partly because the anticuts we have imposed in event selection have different effects in different parts of the Dalitz plot.  Also, interferences have been omitted in evaluating these fractions, which therefore do not add to 100\%.  The last column in the table lists the change in log-likelihood if a particular resonance is dropped from the fit, and the parameters of all other resonances are reoptimized.  This provides information about the relative importance of a resonance which is complementary to that given by its relative intensity. We now briefly discuss results shown in Figs. $3-6$ for individual resonances.

Our procedure for obtaining the best fit values of the 
parameters of a resonance is illustrated in Fig. \ref{f01710}(a,b) with 
the results 
for $f_J(1710)$. The $|-ln{L}|$ is calculated for specific values 
of mass, width, and $J^{PC}$ of $f_0(1710)$.
The results are fitted 
with smooth curves to determine the best fit 
values of $M$ and $\Gamma$ corresponding to the maxima in $|-ln{L}|$.
Fig. \ref{f01710}(b) shows that the $J^{PC}=0^{++}$ solution is clearly 
preferred. 
The corresponding contour plot for  $M/\Gamma$ is shown in Fig. 
\ref{f01710}(a). 
The same procedure was followed for $f_J(1500)$, whose results are 
shown in Fig. \ref{f01710}(c). 
Again  $J^{PC}=0^{++}$ is clearly preferred by $\Delta(-\ln L)=46$.
The results for other resonances are shown 
are shown in Figs. 
\ref{f02020f22150}, \ref{f22300a01450}, and  
\ref{a21700a42240}.
In all but one case the $|-ln{L}|$ differences between 
the different spin choices are large, and the spin assignment 
with the largest  $|-ln{L}|$ (shown with solid points) is preferred; 
the others, (shown with open points) are rejected. 
For $a_J(2040)$ (Fig. 6), although the  $|-ln{L}|$ values for $J^{PC}=2^{++}$ and 
$4^{++}$ are comparable, no maxima are found for either 
$M$ or $\Gamma$ for  $J^{PC}=2^{++}$, and it is also rejected.

\subsection{The Isoscalar States $f_0$ and $f_2$.}

As already mentioned, Fig. \ref{f01710} clearly shows that both $f_0(1500)$ and 
$f_0(1710)$ have $J^{PC}=0^{++}$. 
The $J^{PC}=0^{++}$ assignment is preferred over $2^{++}$ by 
$\Delta ln{L}=58$ for $f_0(1500)$, and $\Delta ln{L}=46$ for 
$f_0(1710)$.
As listed in Table \ref{tab:b}, for $f_0(1500)$ our mass is $\sim30$ MeV smaller than the PDG average, but the width is identical.  For $f_0(1710)$ our mass is $\sim30$ MeV larger, and the width is also $\sim30\%$ larger.
We note that the Crystal Barrel has reported that while they do observe $f_0(1500)$, they do not find 
any evidence for $f_0(1710)$ excitation in the  
$\bar p p \rightarrow \eta\eta\pi^0$ reaction in $\bar p p$ 
annihilation at either 0.9 GeV/c 
\cite{Amsler:qq} or 1.94 GeV/c  \cite{Abele:pf}. On the other hand,   
WA102 \cite{Barberis:2000cd} reports excitation of both   
$f_0(1500)$ and $f_0(1710)$ in central 
production with decay into $\eta\eta$ .
We find that in our $\bar p p \rightarrow \eta\eta\pi^0$ measurement at 5.2 
GeV/c, in the Dalitz plot the ratio of the intensities 
$(f_0(1710)\rightarrow\eta\eta)/(f_0(1500)\rightarrow\eta\eta)\sim 2.2$.
We note that Amsler et al. \cite{Amsler:qq} report this ratio to be $<0.25$ (90\% CL) for production in $p\bar{p}$ annihilation at 0.9 GeV/$c$.
Hopefully, our result provides a new input in the continuing discussion 
about the scalar glueball admixtures in these two states. 


\begin{table*}[!tb]
\newcommand{\m}{\hphantom{$*-$}}
\newcommand{\cc}[1]{\multicolumn{1}{c}{#1}}
\renewcommand{\arraystretch}{1.2} 
\begin{tabular}{@{}l|c|c|c|c|c|r|c}
\hline
\hline
  
& \multicolumn{2}{c|}{Mass(MeV)} & 
& \multicolumn{2}{c|}{Width(MeV)} & Rel. Int. & If removed  \\
\hline
Resonance  &  PDG'04 & Present & $J^{PC}$    &  PDG'04 & Present &   \% & ${\frac{\Delta ln{L}}{\Delta\#par.}}$ \\
\hline

\hline
$f_0(1500)$  
& $1507\pm5$ &$1473\pm5$  & 
$0^{++}$   
& $ 109\pm7$ & $108\pm9$ & $2.4$ & 207/1 \\ 
\hline

$f_0(1710)$  
& $1714\pm 5$ & $1747\pm5$  & 
$0^{++}$ 
& $ 140\pm10$ & $ 188\pm13 $ & $ 5.2$ & 521/3 \\ 
\hline

$f_0(2020)^{*\dagger}$ 
& $1992\pm16$ & $2037\pm8$ & 
$0^{++}$  
& $442\pm60$  & $296\pm17$ & $10.9$ & 295/3 \\ 
\hline

$f_0(2100)^{*\dagger}$  
& $2103\pm7$ & $2105\pm8$ & 
$0^{++}$  
& $206\pm15$  & $236\pm14$ & $7.9$ & 249/3 \\ 
\hline

$f_2(2150)^*$  
& $2156\pm11$ & $ 2170\pm6 $ & 
$2^{++}$  
& $ 167\pm30$ &  $ 182\pm11$ & $ 6.5$ & 209/6 \\ 
\hline

$f_2(2340)^{*\dagger}$  
& $2339\pm55$ & $2350\pm7$ & 
$2^{++}$   
& $319^{+81}_{-69}$ &  $218\pm16$ & $4.8$ & 124/2 \\ 
\hline

\hline

$a_2(1320)$  
& $1318\pm1$ &  $1327\pm2$ & 
$2^{++}$   
& $111\pm2$ &  $128\pm4$ & $ 6.5$ & 2335/14 \\ 
\hline

$a_0(1450)$  
& $1474\pm19$ & $1477\pm10$ &  
$0^{++}$
& $265\pm13$ & $267\pm11$ & $ 3.9$ & 418/5 \\ 
\hline

$a_2(1700)^*$  
& $1732\pm16$ & $ 1702\pm7       $ &  
$2^{++}$ 
& $ 194\pm40$ & $ 417\pm19$ & $20.6$ & 1185/4 \\ 
\hline

$a_4(2040)$  
& $2010\pm12$ & $2004\pm6$ & 
$4^{++}$   
& $ 353\pm40$ & $401\pm16$ & $20.1$ & 738/13 \\ 
\hline
$a_4(2237)^*$  
& $-$ & $2237\pm5$ & 
$4^{++}$  
& $-$ & $291\pm12$ & $20.7$ & 823/4 \\ 
\hline
\end{tabular}\\[2pt]
\caption{Results for  masses, widths and  $J^{PC}$ 
of light quark resonances decaying into $\eta \eta$ and $\eta \pi^0$  
as seen in  $\bar p p \rightarrow \pi^0 \eta \eta $ at 5.2 GeV/c.
All errors are statistical only and correspond to 
$\Delta |-ln{L}|=0.5$ or $\Delta \chi^2=\pm1\sigma$. 
Asterisks$^*$ mark states presently omitted
from meson summary list of PDG'04. Daggers$^{\dagger}$ mark
f-states which have not been seen in $\eta \eta $ before. 
For explanation of the last two columns, see text.
}
\label{tab:b}
\end{table*}

Although the observation of the isoscalars $f_0(2020)$ and $f_0(2100)$
has been reported before, PDG'04 considers them unconfirmed. 
We find $f_0(2020)$ and $f_0(2100)$ with masses 
and widths listed in the Table \ref{tab:b} 
to be strongly populated with nearly equal intensity. 
Once again the $0^{++}$ assignments are preferred for both over 
$2^{++}$ assignments  by $\Delta ln{L}=23$ and 22, respectively.
In a global analysis of the Crystal Barrel data for momenta 900-1900 MeV/c 
Anisovich \it et al. \rm \cite{Anisovich:2000ae} identify these two scalars decaying into 
$\eta\eta$ as having mass/width of $(2005\pm30)/(305\pm50)$ MeV and 
$(2105\pm15)/(200\pm25)$ MeV, which are in almost exact agreement
with ours. We also note that a combined enhancement of these two states 
was reported as X(2100) decaying into $\eta\eta$ by the earlier Fermilab E760 
experiment \cite{Armstrong:1993fh}. Its $J^{PC}$ was not determined.

Although ten $f_2$-states have been reported in our 
mass region by different experiments, the PDG regards only four, 
$f^\prime_2(1525)$, $f_2(2010)$, $f_2(2300)$ and 
$f_2(2340)$ as sufficiently well established to be included in the meson
summary list. None of these except $f^\prime_2(1525)$ have been observed 
to decay into $\eta\eta$. We find no evidence for the excitation of 
$f^\prime_2(1525)$, although it is possible that its resolution 
is rendered impossible due to the presence of nearby stronger 
$f_0(1500)$. 
The $ln{L}$ plot in Fig. \ref{f01710}(c) suggests that 
$f_2(1525)$ population is very small, if not zero.
Similarly, we find no evidence for $f_2(2010)$. 
We confirm the existence of $f_2(2150)$ which has been omitted 
by PDG from its summary list, but which was previously 
observed to decay into $\eta\eta$ by WA102 \cite{Barberis:2000cd}.
Our mass and width are in good agreement with both WA102 and the PDG average.
We also observe clear evidence for $f_2(2340)$. Etkin \it et al. \rm 
\cite{Etkin:1987rj} had claimed this as a doublet, 
$f_2(2300)$ and $f_2(2340)$,  in its $\phi\phi$ decay.
We do not find any evidence for such a close mass doublet.

We note that we do not find any evidence for the tensor glueball 
candidate $\xi(2230)$, or $f_2(2230)$ decaying into $\eta\eta$.
This is consistent with the result of the Crystal Barrel search for the same
\cite{Amsler:2001fh}.

\subsection{The Isovector States $a_0$, $a_2$ and $a_4$.}

Our fit to the $\eta\pi^0$ mass spectrum (see fig. \ref{best}(c)) is admittedly 
poorer compared to the isoscalar state in fig \ref{best}(b). 
It also presents a certain vexing problem concerning the strong 
enhancement observed at mass $\sim 1320$ MeV, which 
is followed by a dip in the vicinity of 1450 MeV.
The $a_2(1320)$ is a well established tensor resonance with PDG mass/width,
$M/\Gamma=(1318.3\pm0.6)/(105-111)$ MeV. It is strongly populated 
in many reactions, and its decays into several channels, including 
$\eta\pi^0$, are well documented \cite{Eidelman:2004wy}. A scalar
$a_0(1450)$ with $M/\Gamma=(1470\pm25)/(265\pm30)$ MeV decaying into 
$\eta\pi^0$ was reported by CB \cite{Amsler:1994pz} in $\bar p p$ 
annihilation at rest, and also in $K_LK^{\pm}$ \cite{Abele:1998qd}.

On the other hand, OBELIX \cite{Bertin:1998sb} has reported a scalar with  
$M/\Gamma=(1290\pm10)/(80\pm5)$ in $\bar p p$ annihilation
decaying into $K^{\mp}K^{0}_S$, along with a nearly factor five weaker excitation of $a_2(1320)$ than what follows from the PDG averages.
We have examined the possible existence of $a_0(1290)$ of OBELIX.
As shown in Fig. \ref{f22300a01450}(b) and Table \ref{tab:b} 
we obtain a good fit to our data with $a_2(1320)$ and $a_0(1450)$, 
with the best fit $M/\Gamma$ 
parameters for both, $a_2$: $M/\Gamma=(1327\pm2)/(128\pm4)$ MeV 
and $a_0$: $M/\Gamma=(1477\pm10)/(267\pm11)$, both in 
excellent agreement with the PDG'04 values.
Paradoxically, an even better fit (not shown) is obtained with $a_0(1290)$ with $M/\Gamma=(1314\pm2)/(133\pm4)$ MeV 
and $a_0(1450)$ with $M/\Gamma=(1404\pm9)/(222\pm18)$ MeV. 
With a 66\% (or $8.3\sigma $) larger width for $a_0(1290)$ 
than that reported by OBELIX ($133\pm4$ MeV versus $80\pm5$ MeV)
it is unlikely that this solution represents their resonance. 
However, this still poses the 
question whether both $a_2(1320)$ and $a_0(1290)$ can coexist. 
We have studied this question at length, considered  
additional interferences, and conclude that if 
both resonances exists, it is not possible to resolve them. 
We list the results for the $J=2$ assignment in Table \ref{tab:b} 
because of the weight of numerous final states, including 
$\eta\pi$, in which $a_2(1320)$ has been observed, and not $a_0(1290)$.
Nevertheless, we must consider the ambiguity between $a_2(1320)$ and 
$a_0(1290)$ as unresolved.

We confirm $a_4(2040)$ with parameters in excellent agreement with those 
in PDG'04. We confirm $a_2(1700)$ but find it to have much larger width than PDG'04. We find a strong resonance, $a_4(2237)$ decaying into $\eta\pi$, with $M/\Gamma=(2237\pm5)/(291\pm15)$ MeV.  A $J^{PC}=4^{++}$ resonance decaying into $\eta\pi$ and $\eta'\pi$ was earlier reported by Anisovich et al. \cite{Anisovich:1999b}.

Additional resonances which were tried in the fit 
were $a_0(980)$, $f_0(1370)$, 
$f_2^{\prime}(1525)$, 
$f_2(1640)$, $f_2(1910)$, $\pi_1(1400)$, $\pi_1(1600)$,  
$f_0(1710-1790)$,  $f_2(1870-1910)$. 
None of these obtain 
significant contributions ($>0.5\%$) to the intensity. 
As mentioned earlier, the resonances near the boundaries of the 
Dalitz Plot, particularly $a_0(980)$ and $f_0(1370)$, suffer from the effect of the 'anticuts'.
No convergence was obtained when the reported hybrids $\pi_1(1400)$ 
and $\pi_1(1600)$ were included in the fits.

The authors wish to thank the E835 Collaboration for allowing us the use 
of their data.
This work was supported by the US Department of Energy.

\end{document}